\documentclass[prl,twocolumn,preprintnumbers,amsmath,amssymb]{revtex4}

\usepackage{graphicx}% Include figure files
\usepackage{dcolumn}% Align table columns on decimal point
\usepackage{bm}% bold math

\begin{document}

\title{Quantum Critical Behavior of the Cluster Glass Phase}

\author{Matthew J. Case}
 \email{case@magnet.fsu.edu}
\author{V. Dobrosavljevi\'{c}}
 \altaffiliation[Also at ]{Department of Physics, FSU}
\affiliation{
National High Magnetic Field Laboratory,
Florida State University, Tallahassee, Florida, 32306
}

\date{\today}

\begin{abstract}
In disordered itinerant magnets with arbitrary symmetry of the
order parameter, the conventional quantum critical point between
the ordered phase and the paramagnetic Fermi-liquid (PMFL) is
destroyed due to the formation of an intervening cluster glass
(CG) phase. In this Letter we discuss the quantum critical
behavior at the CG-PMFL transition for systems with continuous
symmetry. We show that fluctuations due to quantum Griffiths
anomalies induce a first-order transition from the PMFL at T=0,
while at higher temperatures a conventional continuous transition
is restored. This behavior is a generic consequence of enhanced
non-Ohmic dissipation caused by a broad distribution of energy
scales within any quantum Griffiths phase in itinerant systems.
\end{abstract}

\pacs{Valid PACS appear here}
\maketitle

In magnetic systems with disorder, it is always possible to find
isolated, defect-free regions which act as finite-size copies of a
clean, bulk sample.  Fluctuations of these rare regions (or {\it
droplets}) lead to singularities of the free energy in what is
known as the Griffiths phase \cite{Griffiths:1969be}. While these
singularities are extremely weak (essential) singularities in
classical systems \cite{PhysRevB.10.4665}, the situation can
change dramatically near (T=0) quantum phase transitions
\cite{Fisher:1992qf,
fisher:6411,Motrunich:2000zt,PhysRevB.66.174433,vojta:107202,
vojta:045438,dobrosavljevic:187203}. Recently, there has been
significant progress made in understanding the quantum Griffiths
phase by focusing on the dynamics of the rare regions
\cite{0305-4470-39-22-R01}.  By considering the rare regions as
independent droplets described by a local order parameter, an
elegant classification of quantum Griffiths phenomena was proposed
based only on general symmetry principles \cite{vojta:045438}.

In the independent droplet picture, the system is assumed to be correlated
over the length scale of the individual droplets, each of which has
dynamics in the imaginary time direction related to the energy cost of
coherently flipping a large volume of spins.  The local droplet degrees of
freedom therefore map to classical, one-dimensional ferromagnetic spin chains
in the imaginary-time direction,
with their dynamics corresponding to interactions in this single dimension.
The critical behavior of the bulk system can then be understood through the
behavior of the independent spin chains which depends only on their dynamics
and the spin symmetry \cite{Kosterlitz:1976yv}.  For example,
in the case of itinerant Ising magnets, independent spin chains are effectively
above their lower critical dimension and undergo
a phase transition in the universality class of the Kosterlitz-Thouless
transition \cite{Kosterlitz:1976yv}.  Thus, if sufficiently
large, these droplets freeze over, leading to their eventual ordering at
low enough temperature.  Therefore, at zero temperature, the system is always
in its ordered phase and the quantum phase
transition present in the clean magnet becomes destroyed or {\it rounded}
due to the presence of disorder
\cite{vojta:107202}. In contrast, independent droplets in
itinerant Heisenberg magnets are exactly at their lower critical
dimension \cite{Kosterlitz:1976yv}, corresponding to weaker dissipation.
Within the
independent droplet picture, this behavior leads to quantum
Griffiths effects, and the quantum phase transition to the
magnetically ordered phase is conjectured to be in the
universality class of the infinite randomness fixed point (IRFP)
\cite{vojta:045438}. These general issues have attracted much
interest in recent years due to the possible role of such quantum
Griffiths phase phenomena as a driving force for the
disorder-driven non-Fermi liquid behavior in correlated electron
systems
(for a recent review see Ref.~\onlinecite{0034-4885-68-10-R02}).

In realistic systems, however, the droplets are not independent
and interactions must be considered, providing a qualitatively
different mechanism for dissipation. Such interaction effects were
recently \cite{dobrosavljevic:187203} shown to represent a
singular perturbation within any quantum Griffiths phase, leading
to enhanced non-Ohmic dissipation. This mechanism leads to the
freezing of sufficiently large droplets and the generic emergence
of a cluster glass (CG) phase intervening between the uniformly
ordered phase and the paramagnetic Fermi liquid (PMFL).  (For a schematic
phase diagram of an itinerant magnet with and without interactions,
we refer the reader to Ref.~\onlinecite{dobrosavljevic:187203}.)

In this Letter, we examine the nature of the quantum phase
transition to the cluster glass phase.  To study the critical
behavior at the quantum CG-PMFL transition, we focus on the
continuous spin symmetry case for which the interaction effects
produce a sharp phase transition at $T=0$.
By studying the transition both from the CG and from the PMFL,
we show
that a region of coexistence of these two phases exists, indicating that the
$T=0$ quantum phase transition is first-order.  At higher
temperatures, a tricritical point is found above which a
conventional second-order transition is restored.
In contrast, we show that a
collection of identical droplets displays a conventional
(continuous) phase transition even down to the lowest
temperatures.  From this we conclude that fluctuations due to Griffiths phase
anomalies are responsible for driving the transition first-order.

{\it Self-consistency conditions.}---
Our approach here mirrors that of previous works \cite{vojta:045438,
dobrosavljevic:187203}.  In the presence of
disorder, defect-free regions can form with near uniform ordering while the
bulk system is in the non-ordered phase.  These rare regions (droplets) can
be expressed as a fluctuating order parameter field in the spirit of
Ginzburg-Landau with effective action
$S = \sum_{i}{\cal S}_{{\rm L}, i}
+\sum_{ij}{\cal S}_{{\rm I},ij}$.  For the case of an itinerant
antiferromagnet \cite{Hertz:1976kl}, the local action (for a single droplet) is
\begin{eqnarray}
{\cal S}_{{\rm L},i}&=&\sum_{\omega_n}\phi_i(\omega_n)(r_i+|\omega_n|)
\phi_i(-\omega_n)\\ \nonumber
&&+\frac{u}{2N}\int_0^{\beta}{\rm d}\tau\phi^4_i(\tau)
\end{eqnarray}
where $\phi_i$ is the $N$-component order parameter field for the $i^{\rm th}$
droplet and $r_i$ is its bare mass, selected from the distribution
${\cal P}(r_i)$.  This local action has been studied previously and is well
understood \cite{vojta:045438}.  The
novelty arises when we consider interactions between droplets, mediated by the
weakly correlated metal in the bulk.  These take the form of the RKKY
interaction
\begin{equation}
{\cal S}_{{\rm I}, ij}=\frac{J_{ij}}{(R_{ij})^d}
\int_0^{\beta}{\rm d}\tau\phi_i(\tau)\phi_j(\tau),
\end{equation}
with $J_{ij}$ random, of zero mean and having variance
$\left<J_{ij}^2\right>=J^2$.
Very recent work has suggested that
even infinitesimally weak RKKY interactions destabilize the
Griffiths phase found in the non-interacting droplet theory,
leading to the cluster glass phase \cite{dobrosavljevic:187203}.
These results demonstrated that the nontrivial physics of
non-Ohmic dissipation persists even when the problem is solved at
the saddle-point level, which is formally justified in the
large-$N$ limit of the model \cite{zinn}.

We proceed in the usual way by averaging over disorder using the
replica method and decoupling the resultant quartic term by
introducing an auxiliary Hubbard-Stratonovich field
\cite{Binder:1986lm, dobrosavljevic:187203}.
This new field describes the dynamics of
spin fluctuations caused by the long-ranged RKKY interactions, and
contributes an additional dissipative term to the action. Its
saddle-point value is \cite{dobrosavljevic:187203}
\begin{equation}
\label{chicond}
\overline{\chi(\omega_n)} = \frac12\int{\rm d}r_i\frac{{\cal P}(r_i)}
{r_i+\lambda_i+|\omega_n|-\tilde{g}\overline{\chi(\omega_n)}}
\end{equation}
where $\tilde{g}\equiv J^2\sum_i(R_{ij})^{-2d}$ is the RKKY coupling.
$\lambda_i$ is a Hubbard-Stratonovich field decoupling the quartic
interaction in the single-site action and satisfies the local
self-consistency condition \cite{zinn}
\begin{equation}
\label{lamcond}
\lambda_i = \frac{uT}{2}\sum_{\omega_n}\frac1
{r_i+\lambda_i+|\omega_n|-\tilde{g}\overline{\chi(\omega_n)}}.
\end{equation}
Our original action now becomes a single-site problem with effective action
\begin{equation}
S_{\rm
eff}=\sum_{i,\omega_n}\phi_i(\omega_n)(r_i+\lambda_i+|\omega_n|-\tilde{g}
\overline{\chi(\omega_n)})\phi_i(\omega_n).
\end{equation}
Physically, this corresponds to independent droplets in the presence of
additional
dissipation arising due to interactions with all other droplet degrees of
freedom and represented by the $\omega$ dependence of the local spin
susceptibility
$\overline{\chi(\omega_n)}$.

{\it Quantum criticality for uniform droplets.}--- An interesting
limit of this problem is revealed if all the droplets are assumed
to have the same size ${\cal P}(r_i)=\delta(r_i-\hat{r})$ for all
sites $i$. In this case, the coupled set of equations
\eqref{chicond} and \eqref{lamcond} are identical to those
describing a metallic spin glass \cite{sachdev} and, at large
enough $\tilde{g}$, we expect a zero temperature transition from a
paramagnet to a spin glass. The self-consistency condition
\eqref{chicond} for $\chi(\omega)$ reduces to a simple algebraic
equation and is easily solved:
\begin{eqnarray}
\label{chisoln}
\chi_{\rm u}(\omega_n)&=&\frac{1}{2\tilde{g}}\bigg(
\hat{r}+\hat{\lambda}+|\omega_n|\\ \nonumber
&&\pm\sqrt{(\hat{r}+\hat{\lambda}+|\omega_n|)^2-2\tilde{g}}\bigg)
\end{eqnarray}
where $\hat{\lambda}\equiv\lambda(\hat{r})$.
Clearly, the paramagnetic solution becomes unstable when
$(\hat{r}+\hat{\lambda})=\sqrt{2\tilde{g}}$ and we define an energy scale 
\cite{PhysRevLett.70.4011}
$\Delta=(\hat{r}+\hat{\lambda}-\sqrt{2\tilde{g}})$.  The phase boundary can be
determined by setting $\Delta=0$ and solving the self-consistency condition
\eqref{lamcond} for $\hat{r}(T)$.

The phase for $\Delta>0$ ($\hat{r}>\hat{r}_{\rm c}$) is characterized by the
frequency dependence of $\chi_{\rm u}(\omega_n)$.
From the solution \eqref{chisoln}, we find three separate regimes,
\begin{equation}
\chi_{\rm u}(0)-\chi_{\rm u}(\omega)\sim
\left\{
\begin{array}{cc}
\omega&,\omega<\Delta\ll\sqrt{2\tilde{g}}\\
\sqrt{\omega}&,\Delta<\omega\ll\sqrt{2\tilde{g}}\\
\chi_{\rm u}(0)-\omega^{-1}&,\omega\gg\sqrt{2\tilde{g}}
\end{array}
\right..
\end{equation}
The linear, low-$\omega$ behavior of $\chi_{\rm u}$ is characteristic of a
Fermi liquid, while, right at the transition, $\chi_{\rm u}\sim\sqrt{\omega}$,
characteristic of non-Fermi liquid behavior.

The transition at $T=0$ can also be studied from the magnetically ordered
side.  In this case, the quantum critical point is identified when the
mean-field stability criterion for the glass phase vanishes \cite{Pastor:4642}
i.e. $\widetilde{\lambda}_{\rm SG}\equiv1-\sqrt{2\tilde{g}}\chi(0)=0$.
Again, using the solution \eqref{chisoln} for $\chi$, we find
$\widetilde{\lambda}_{\rm SG}\sim\sqrt{\Delta}$
which vanishes at $\Delta=0$.

Thus, examining the transition through the instability of either the
paramagnetic Fermi liquid or the spin glass phase yields a quantum critical
point at $\Delta=0$, consistent with a conventional second-order transition in
this limit of the problem.

In the following
we consider the model with a distribution of droplet sizes, and
show that the two approaches do not coincide at the transition,
revealing the singular effects of droplet fluctuations.

{\it Griffiths phase behavior of distributed droplets.}--- For
dilute impurities, the defect-free regions assume a Poisson
distribution, which can be written in terms of the local coupling
constant as
\begin{equation}
{\cal P}(r_i)=\frac{2\pi\alpha}{u}e^{2\pi\alpha(r_i-\hat{r})/u},\hspace{1em}
r_i\le\hat{r}<0.
\end{equation}
The offset $\hat{r}$ is used to tune through the transition,
$\alpha$ is related to the droplet density $\rho$ by $\alpha=\rho
u/2\pi\delta$ and $\delta<0$ is the coupling constant of the clean
magnet.  In this way, the distribution is smoothly connected to
the uniform-$r_i$ limit as $\alpha\rightarrow\infty$.  We can
therefore reasonably apply the uniform solution \eqref{chisoln} as
a zeroth order approximation and iterate equations \eqref{chicond}
and \eqref{lamcond} until self-consistency is achieved.  By
accounting for droplets of all sizes, we are also including their
fluctuations which are associated with the Griffiths behavior.

The instability of the Griffiths phase at $T=0$ is already
apparent at one iteration loop (1IL). Physically, this corresponds to
the regime where very large droplets are very dilute, but the
typical droplets (in their local environment) are approaching quantum
criticality of the uniform droplet model.
Integrating expression \eqref{lamcond} and identifying the local
droplet energy
$\epsilon_i=\hat{r}+r_i+\lambda_i-\tilde{g}\chi(0)$, we find the
relation between droplet energy and coupling strength
$\epsilon_i\sim\exp(2\pi f^{-1}r_i/u)$ where we have defined
\begin{equation}
\label{fdef}
f\equiv\frac{2\sqrt{\Delta^2+2\sqrt{2\tilde{g}}\Delta}}
{\Delta+\sqrt{2\tilde{g}}+\sqrt{\Delta^2+2\sqrt{2\tilde{g}}\Delta}}.
\end{equation}
We can now switch from integration over local coupling constants
to integration over energies via the replacement $\int{\cal
P}(r_i)(\cdots){\rm d}r_i\rightarrow\int{\cal
P}(\epsilon_i)(\cdots){\rm d}\epsilon_i$ where the distribution of
droplet energies is a {\it power-law} ${\cal
P}(\epsilon)\sim\epsilon^{\alpha'-1}$, $\alpha'\equiv f\alpha$.
From Eq. (11) we conclude that the ``renormalized Griffiths
exponent'' $\alpha '$ decreases as the critical point is
approached. Scaling arguments \cite{vojta:045438} show that such a
distribution leads to strong Griffiths behavior in the isolated
droplet case for $\alpha'<1$.

For the case of interacting droplets, however, integration of
expression \eqref{chicond} produces
$\overline{\chi(\omega)}-\overline{\chi(0)}\sim-|\omega|^{\alpha'-1}+
{\cal O}(|\omega|)$ and non-Ohmic dissipation obtains for
$\alpha'<2$. As argued previously \cite{dobrosavljevic:187203},
this qualitatively changes the dynamics of the droplets which then
find themselves above their lower critical dimension. Sufficiently
large droplets ($r_i>r_c$) can therefore order (freeze) in
imaginary time before the Griffiths phase is reached, and this new
phase has been dubbed the ``cluster glass''.

{\it IRFP of isolated droplets at quantum criticality.}-- It
should be pointed out that our 1IL expression for $\alpha'$
underscores the close analogy between the present study and the
physics of infinite-randomness fixed points (IRFP).  Most
remarkably, from Eq. (11) we conclude that, within such 1IL
calculation, the ``renormalized Griffiths exponent'' $\alpha'$
vanishes precisely at the uniform droplet quantum critical point
as
\begin{equation}
\alpha ' \sim \Delta^{1/2} .
\end{equation}
This result should be contrasted with the theory of noninteracting
droplets \cite{vojta:045438} where the ``bare'' Griffiths
exponent $\alpha$ is a non-universal function of parameters.

The dynamical critical exponent governing an IRFP is commonly
defined \cite{PhysRevB.56.11691}
as $z'=d/\alpha'$ ($d$ is the spatial dimensionality),
diverging right at the transition.
Equation \eqref{fdef} provides us with an explicit expression for
$z'$ which captures both the quantum Griffiths phase and the
behavior reminiscent of an IRFP at $\Delta=0$.  Of course, this
expression is not fully self-consistent, and will not be
appropriate at the lowest temperatures where the system
freezes into the cluster glass state. However, quantum Griffiths
and IRFP behavior obtained from the 1IL approximation is
expected to hold at temperatures above the cluster glass
transition, as anticipated previously \cite{vojta:045438}.

{\it Fluctuation induced first-order transition.}---
In the paramagnetic phase, it is clear that the leading order behavior of
$\overline{\chi(\omega)}$ is linear in $\omega$ and it is not difficult to
show that the
slope $m_{\chi}=\partial\overline{\chi}/\partial\omega|_{\omega=0}$ determines
the renormalized exponent $\alpha'=f\alpha$ through the factor
$f=(1-\tilde{g}m_{\chi})^{-1}$, as can be checked at the 1IL level.
This relationship can be further illuminated
by calculating $m_{\chi}$ self-consistently from equation \eqref{chicond}:
\begin{equation}
m_{\chi}=-2(1-\tilde{g}m_{\chi})\overline{\chi^2(0)}.
\end{equation}
Solving for the slope, we find the relationship
\begin{equation}
\label{mX2rel}
f=(1-\tilde{g}m_{\chi})^{-1}=(1-2\tilde{g}\overline{\chi^2(0)}).
\end{equation}
What is striking about this result is that the term in parentheses on the
right-hand side of \eqref{mX2rel} is the exact analogue of the spin glass
stability criterion $\widetilde{\lambda}_{\rm SG}$ for droplets with
distributed site-energies \cite{Pastor:4642}; in this case, the
stability criterion is $\lambda_{\rm SG}=1-2\tilde{g}\overline{\chi^2(0)}$.
The fully self-consistent value of $\alpha'$
is then $\alpha'=f\alpha=\lambda_{\rm SG}\alpha$, implying
that $\alpha'=0$ at the
transition determined by the stability of the CG phase.
On the other hand, approaching from the PMFL side, the transition
is still given by
$\alpha'=2$ where non-Ohmic dissipation obtains.
So, unlike the uniform droplet case, the stability criteria for the PMFL and
the CG do not coincide at a unique point, implying that the quantum phase
transition in the distributed droplet case is not a conventional second-order
transition.

To gain further insight into the nature of the transition, we recast the
problem as an eigenvalue analysis of the constitutive free energy for which
the solution to \eqref{chicond} is a minimum.  The relevant contribution to
the free energy can be expressed near the minimum as
\begin{equation}
{\cal F}_{\chi} = \int_{\omega,\omega'}
(\chi(\omega)-\chi_0(\omega))\Gamma_{\chi}(\omega,\omega')
(\chi(\omega')-\chi_0(\omega'))
\end{equation}
so that the eigenvalue determining the stability of the solution $\chi_0$
satisfies
\begin{equation}
\label{lamchidef}
\lambda_{\chi}\le\frac{\int_{\omega,\omega'}
(\chi(\omega)-\chi_0(\omega))\Gamma_{\chi}(\omega,\omega')
(\chi(\omega')-\chi_0(\omega'))}
{\int_{\omega}(\chi(\omega)-\chi_0(\omega))^2}.
\end{equation}
To implement this iteratively, we can approximate the difference
$(\chi-\chi_0)\approx(\chi^{(n-1)}-\chi^{(n)})$, where $\chi^{(n)}$ is the
value of $\chi$ at the $n$th iteration step.  To determine how $\Gamma_{\chi}$
operates on $\chi^{(n)}$, note that the free energy functional is constructed
such that $\delta{\cal F}/\delta\chi=0$ yields the self-consistency condition
(\ref{chicond}) for $\chi$.  Thus,
$\int{\rm d}\omega'\Gamma_\chi(\omega,\omega')\chi^{(n)}(\omega')
=\chi^{(n)}(\omega)-\chi^{(n+1)}(\omega)$.
We can now rewrite the defining
relation \eqref{lamchidef} for the eigenvalue valid at each iteration step
\begin{equation}
\lambda_{\chi}^{(n)}=1-
\frac{\int{\rm d}\omega(\chi^{(n-1)}-\chi^{(n)})(\chi^{(n)}-\chi^{(n+1)})}
{\int{\rm d}\omega(\chi^{(n-1)}-\chi^{(n)})^2}
\end{equation}
and proceed by iterating numerically.  The instability of the PMFL is then
given by $\lambda_\chi=0$ at $n\rightarrow\infty$.

\begin{figure}[t]
{\centering{\includegraphics[scale=0.76]{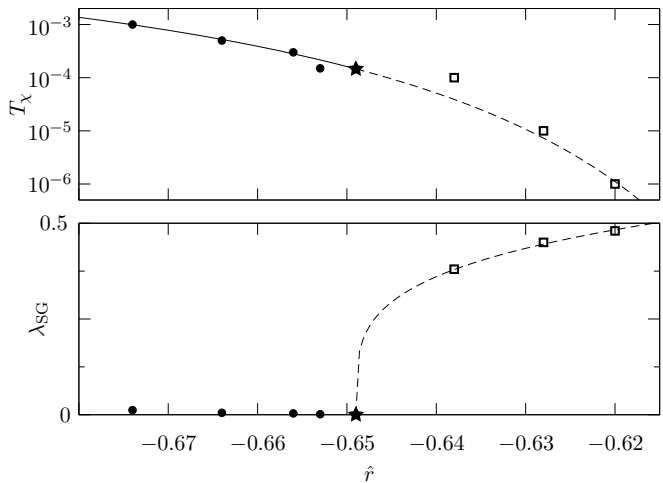}}\par}
\caption{{\it Upper panel}: The boundary corresponding to $\lambda_\chi=0$ is
shown for finite temperatures.  The line is a fit to the data as discussed in
the text.
{\it Lower panel}: The stability criterion $\lambda_{\rm SG}$
calculated at $T_\chi$.  The line is a guide to the eye.  In both panels, open
squares correspond to first-order transitions, full circles to
second-order transitions.  The tricritical point is indicated by the solid
star.} \label{pb}
\end{figure}

The results of this procedure at finite temperature are shown in the upper
panel of Figure \ref{pb}.  Plotted in both full and empty symbols is the
temperature $T_\chi$ at which
$\lambda_\chi=0$ for given values of the tuning parameter $\hat{r}$.
Plotted in the lower panel is the value of $\lambda_{\rm SG}$ at
$T_\chi(\hat{r})$.
We can see that $\lambda_{\rm
SG}>\lambda_{\chi}=0$ for $-0.65\alt\hat{r}\le\hat{r}_{\rm c}(T=0)$ which
indicates that there is a region of coexistence of the CG and PMFL phases
where $\lambda_\chi$ and $\lambda_{\rm SG}$ are both finite and positive (i.e.
both solutions are stable).
The transition for $\hat{r}>-0.65$ is therefore of first-order, so the
open squares represent the spinodal line; the other spinodal
corresponds to $\lambda_{\rm SG}=0, \lambda_\chi>0$.
For $\hat{r}<-0.65$, the two criteria merge and the usual continuous transition
is restored.  The boundary given by $\lambda_\chi=0$ in this regime
is then the true phase boundary which we can fit using
the form $\log(T_{\rm c})\sim(\hat{r}-\hat{r_{\rm c}})^{-1}$, as shown by
the full line in the upper panel, reflecting the
fact that the critical temperature is proportional to the number of frozen
droplets \cite{dobrosavljevic:187203}.
Thus, open squares represent a discontinuous
transition for $\hat{r}>-0.65$ while the full circles represent a
continuous transition for $\hat{r}<-0.65$; the tricritical point at
$\hat{r}=-0.65$ is shown by the solid star.

In summary, we examined quantum criticality of the cluster glass
phase which typically emerges as disorder is introduced in
itinerant magnets. We have shown how coupling between
droplets enhances the quantum Griffiths effects, leading to
singularly enhanced dissipation in the vicinity of the quantum
critical point. As a result, any system with widely distributed
droplet sizes cannot display conventional quantum criticality, but
instead features a first-order transition at
$T=0$. In contrast, if the effects of the largest droplets
are eliminated, conventional criticality is restored, as we
found in a related model with a bounded distribution of droplet
sizes. Similar behavior is also found at sufficiently high
temperatures, where thermal fluctuations impose a cutoff on the
Griffiths phase anomalies. This implies that fluctuations due to these Griffiths
phase anomalies are responsible for driving the transition first-order.
Our results could be of direct
relevance to experiments on disordered superconductors where
singular corrections due to rare regions have recently been
demonstrated \cite{Vafek:2006fk}.

We would like to acknowledge fruitful discussions with Andrey
Chubukov, Eduardo Miranda, Joerg Schmalian, Oskar Vafek, and
Thomas Vojta. This work was supported through the NSF Grant No.
DMR-0542026 (V. D.) and the National High Magnetic Field
Laboratory (M.J.C.). We also thank the Aspen Center for Physics,
where part of this work was carried out.

%\clearpage


\begin{thebibliography}{20}
\expandafter\ifx\csname natexlab\endcsname\relax\def\natexlab#1{#1}\fi
\expandafter\ifx\csname bibnamefont\endcsname\relax
  \def\bibnamefont#1{#1}\fi
\expandafter\ifx\csname bibfnamefont\endcsname\relax
  \def\bibfnamefont#1{#1}\fi
\expandafter\ifx\csname citenamefont\endcsname\relax
  \def\citenamefont#1{#1}\fi
\expandafter\ifx\csname url\endcsname\relax
  \def\url#1{\texttt{#1}}\fi
\expandafter\ifx\csname urlprefix\endcsname\relax\def\urlprefix{URL }\fi
\providecommand{\bibinfo}[2]{#2}
\providecommand{\eprint}[2][]{\url{#2}}

\bibitem[{\citenamefont{Griffiths}(1969)}]{Griffiths:1969be}
\bibinfo{author}{\bibfnamefont{R.~B.} \bibnamefont{Griffiths}},
  \bibinfo{journal}{Phys. Rev. Lett.} \textbf{\bibinfo{volume}{23}},
  \bibinfo{pages}{17} (\bibinfo{year}{1969}).

\bibitem[{\citenamefont{Wortis}(1974)}]{PhysRevB.10.4665}
\bibinfo{author}{\bibfnamefont{M.}~\bibnamefont{Wortis}},
  \bibinfo{journal}{Phys. Rev. B} \textbf{\bibinfo{volume}{10}},
  \bibinfo{pages}{4665} (\bibinfo{year}{1974}).

\bibitem[{\citenamefont{Fisher}(1992)}]{Fisher:1992qf}
\bibinfo{author}{\bibfnamefont{D.~S.} \bibnamefont{Fisher}},
  \bibinfo{journal}{Phys. Rev. Lett.} \textbf{\bibinfo{volume}{69}},
  \bibinfo{pages}{534} (\bibinfo{year}{1992}).

\bibitem[{\citenamefont{Fisher}(1995)}]{fisher:6411}
\bibinfo{author}{\bibfnamefont{D.~S.} \bibnamefont{Fisher}},
  \bibinfo{journal}{Phys. Rev. B} \textbf{\bibinfo{volume}{51}},
  \bibinfo{pages}{6411} (\bibinfo{year}{1995}).

\bibitem[{\citenamefont{Motrunich et~al.}(2000)\citenamefont{Motrunich, Mau,
  Huse, and Fisher}}]{Motrunich:2000zt}
\bibinfo{author}{\bibfnamefont{O.}~\bibnamefont{Motrunich}}
  \textit{et~al.},
  %\bibinfo{author}{\bibfnamefont{S.-C.} \bibnamefont{Mau}},
  %\bibinfo{author}{\bibfnamefont{D.~A.} \bibnamefont{Huse}}, \bibnamefont{and}
  %\bibinfo{author}{\bibfnamefont{D.~S.} \bibnamefont{Fisher}},
  \bibinfo{journal}{Phys. Rev. B} \textbf{\bibinfo{volume}{61}},
  \bibinfo{pages}{1160} (\bibinfo{year}{2000}).

\bibitem[{\citenamefont{Millis et~al.}(2002)\citenamefont{Millis, Morr, and
  Schmalian}}]{PhysRevB.66.174433}
\bibinfo{author}{\bibfnamefont{A.~J.} \bibnamefont{Millis}}
  \textit{et~al.},
  %\bibinfo{author}{\bibfnamefont{D.~K.} \bibnamefont{Morr}}, \bibnamefont{and}
  %\bibinfo{author}{\bibfnamefont{J.}~\bibnamefont{Schmalian}},
  \bibinfo{journal}{Phys. Rev. B} \textbf{\bibinfo{volume}{66}},
  \bibinfo{pages}{174433} (\bibinfo{year}{2002}).

\bibitem[{\citenamefont{Vojta}(2003)}]{vojta:107202}
\bibinfo{author}{\bibfnamefont{T.}~\bibnamefont{Vojta}},
  \bibinfo{journal}{Phys. Rev. Lett.} \textbf{\bibinfo{volume}{90}},
  \bibinfo{pages}{107202} (\bibinfo{year}{2003}).

\bibitem[{\citenamefont{Vojta and Schmalian}(2005)}]{vojta:045438}
\bibinfo{author}{\bibfnamefont{T.}~\bibnamefont{Vojta}} \bibnamefont{and}
  \bibinfo{author}{\bibfnamefont{J.}~\bibnamefont{Schmalian}},
  \bibinfo{journal}{Phys. Rev. B} \textbf{\bibinfo{volume}{72}},
  \bibinfo{pages}{045438} (\bibinfo{year}{2005}).

\bibitem[{\citenamefont{Dobrosavljevi\'c and
  Miranda}(2005)}]{dobrosavljevic:187203}
\bibinfo{author}{\bibfnamefont{V.}~\bibnamefont{Dobrosavljevi\'c}}
  \bibnamefont{and} \bibinfo{author}{\bibfnamefont{E.}~\bibnamefont{Miranda}},
  \bibinfo{journal}{Phys. Rev. Lett.} \textbf{\bibinfo{volume}{94}},
  \bibinfo{pages}{187203} (\bibinfo{year}{2005}).

\bibitem[{\citenamefont{Vojta}(2006)}]{0305-4470-39-22-R01}
\bibinfo{author}{\bibfnamefont{T.}~\bibnamefont{Vojta}}, \bibinfo{journal}{J.
  Phys. A} \textbf{\bibinfo{volume}{39}}, \bibinfo{pages}{R143}
  (\bibinfo{year}{2006}).

\bibitem[{\citenamefont{Kosterlitz}(1976)}]{Kosterlitz:1976yv}
\bibinfo{author}{\bibfnamefont{J.~M.} \bibnamefont{Kosterlitz}},
  \bibinfo{journal}{Phys. Rev. Lett.} \textbf{\bibinfo{volume}{37}},
  \bibinfo{pages}{1577} (\bibinfo{year}{1976}).

\bibitem[{\citenamefont{Miranda and
  Dobrosavljevi\'c}(2005)}]{0034-4885-68-10-R02}
\bibinfo{author}{\bibfnamefont{E.}~\bibnamefont{Miranda}} \bibnamefont{and}
  \bibinfo{author}{\bibfnamefont{V.}~\bibnamefont{Dobrosavljevi\'c}},
  \bibinfo{journal}{Rep. Prog. Phys.} \textbf{\bibinfo{volume}{68}},
  \bibinfo{pages}{2337} (\bibinfo{year}{2005}).

\bibitem[{\citenamefont{Hertz}(1976)}]{Hertz:1976kl}
\bibinfo{author}{\bibfnamefont{J.~A.} \bibnamefont{Hertz}},
  \bibinfo{journal}{Phys. Rev. B} \textbf{\bibinfo{volume}{14}},
  \bibinfo{pages}{1165} (\bibinfo{year}{1976}).

\bibitem[{\citenamefont{Zinn-Justin}(1996)}]{zinn}
\bibinfo{author}{\bibfnamefont{J.}~\bibnamefont{Zinn-Justin}},
  \emph{\bibinfo{title}{Quantum Field Theory and Critical Phenomena}}
  (\bibinfo{publisher}{Cambridge University Press}, \bibinfo{year}{1996}).

\bibitem[{\citenamefont{Binder and Young}(1986)}]{Binder:1986lm}
\bibinfo{author}{\bibfnamefont{K.}~\bibnamefont{Binder}} \bibnamefont{and}
  \bibinfo{author}{\bibfnamefont{A.~P.} \bibnamefont{Young}},
  \bibinfo{journal}{Rev. Mod. Phys.} \textbf{\bibinfo{volume}{58}},
  \bibinfo{pages}{801} (\bibinfo{year}{1986}).

\bibitem[{\citenamefont{Sachdev}(1999)}]{sachdev}
\bibinfo{author}{\bibfnamefont{S.}~\bibnamefont{Sachdev}},
  \emph{\bibinfo{title}{Quantum Phase Transitions}}
  (\bibinfo{publisher}{Cambridge University Press}, \bibinfo{year}{1999}).

\bibitem[{\citenamefont{Ye et~al.}(1993)\citenamefont{Ye, Sachdev, and
  Read}}]{PhysRevLett.70.4011}
\bibinfo{author}{\bibfnamefont{J.}~\bibnamefont{Ye}}
  \textit{et~al.},
  %\bibinfo{author}{\bibfnamefont{S.}~\bibnamefont{Sachdev}}, \bibnamefont{and}
  %\bibinfo{author}{\bibfnamefont{N.}~\bibnamefont{Read}},
  \bibinfo{journal}{Phys. Rev. Lett.} \textbf{\bibinfo{volume}{70}},
  \bibinfo{pages}{4011} (\bibinfo{year}{1993}).

\bibitem[{\citenamefont{Pastor and Dobrosavljevi\'c}(1999)}]{Pastor:4642}
\bibinfo{author}{\bibfnamefont{A.~A.} \bibnamefont{Pastor}} \bibnamefont{and}
  \bibinfo{author}{\bibfnamefont{V.}~\bibnamefont{Dobrosavljevi\'c}},
  \bibinfo{journal}{Phys. Rev. Lett.} \textbf{\bibinfo{volume}{83}},
  \bibinfo{pages}{4642} (\bibinfo{year}{1999}).

\bibitem[{\citenamefont{Young}(1997)}]{PhysRevB.56.11691}
\bibinfo{author}{\bibfnamefont{A.~P.} \bibnamefont{Young}},
  \bibinfo{journal}{Phys. Rev. B} \textbf{\bibinfo{volume}{56}},
  \bibinfo{pages}{11691} (\bibinfo{year}{1997}).

\bibitem[{\citenamefont{Vafek et~al.}(unpublished)\citenamefont{Vafek, Beasley,
  and Kivelson}}]{Vafek:2006fk}
\bibinfo{author}{\bibfnamefont{O.}~\bibnamefont{Vafek}}
  \textit{et~al.},
  %\bibinfo{author}{\bibfnamefont{M.~R.} \bibnamefont{Beasley}},
  %\bibnamefont{and} \bibinfo{author}{\bibfnamefont{S.~A.}
  %\bibnamefont{Kivelson}},
  \bibinfo{journal}{cond-mat/0505688}
  (\bibinfo{year}{unpublished}).

\end{thebibliography}
\end{document}